\documentclass{vospwks2007}
\usepackage{graphicx}

\title{3D Spectroscopy in the Virtual Observatory: Current Status}
\author[1,2,3]{Igor Chilingarian}
\author[4]{Fran\c cois Bonnarel}
\author[5]{Mireille Louys}
\author[3]{Ivan Zolotukhin}
\author[6]{Fr\'ed\'eric Royer}
\author[6]{Isabelle J\'egouzo}
\author[2]{Pierre Le Sidaner}
\author[4]{Pierre Fernique}
\author[4]{Thomas Boch}
\affil[1]{LERMA Observatoire de Paris, France}
\affil[2]{VO-Paris Data Centre, France}
\affil[3]{Sternberg Astronomical Institute, Russia}
\affil[4]{CDS Observatoire de Strasbourg, France}
\affil[5]{LSIIT, ULP, Strasbourg, France}
\affil[6]{GEPI Observatoire de Paris, France}

\begin{document}


\maketitle

\begin{abstract}
Three cornerstones for the 3D data support in the Virtual Observatory are:
(1) data model to describe them, (2) data access services providing access
to fully-reduced datasets, and (3) client applications which can deal with
3D data. Presently all these components became available in the VO. We
demonstrate an application of the IVOA Characterisation data model to
description of IFU and Fabry-Perot datasets. Two services providing SSA-like
access to 3D-spectral data and Characterisation metadata have been
implemented by us: ASPID-SR at SAO RAS for accessing IFU and Fabry-Perot
data from the Russian 6-m telescope, and the Giraffe Archive at the VO Paris
portal for the VLT FLAMES-Giraffe datasets. We have implemented VO Paris
Euro3D Client, handling Euro3D FITS format, that interacts with CDS Aladin
and ESA VOSpec using PLASTIC to display spatial and spectral cutouts of 3D
datasets. Though the prototype we are presenting is yet rather simple, it
demonstrates how 3D spectroscopic data can be fully integrated into the VO
infrastructure.
\end{abstract}

\section{Introduction}
Integral field (or 3D) spectroscopy is a modern technique in astrophysical
observing that was proposed by Georges Court\'es in the late 60's. The idea
is to get a spectrum for every point in the field of view of a spectrograph.
Several instrumental approaches in the optical domain (as well as NIR and
near-UV) exist: scanning Fabry-Perot interferometry, image slicing and
transforming two-dimensional field of view into a slit using Integral-Field
Unit (see review in P\'econtal-Rousset et al., 2004 for a description of
different image slicing techniques).

At present, nearly all large telescopes in the world are equipped with 3D
spectroscopic devices, and rapidly growing volume of data produced by them
pose a number of questions regarding the data discovery and retrieval. In
this paper we demonstrate how 3D data are handled in a framework of the
International Virtual Observatory.

All 3D spectroscopic observations result in datasets having both spatial and
spectral information. They are usually referred as ``datacubes'', though 
sometimes (in case of IFU) they are not regularly gridded in spatial 
dimensions.

There are three cornerstones for the 3D data support in the Virtual 
Observatory:
\begin{enumerate}
\item data model -- an abstract, self-sufficient and standardised
  description of the data
\item data access services -- archives, providing access to fully reduced
  science-ready datasets
\item client applications -- data-model aware software that is able to 
  search, retrieve, and display 3D data, as well as to give a possibility
  for sophisticated scientific data analysis
\end{enumerate}

All these blocks became available, and we will review them in the 
forthcoming sections.

\section{Data Model}
An abstract, self-sufficient and standardised description of the
astronomical data is known as a data model. Such a description is
constructed in a way to become sufficient for any sort of data processing
and analysis. The Data Modeling working group (DM WG) of the International Virtual
Observatory Alliance (IVOA) is responsible for definition of data models for
different types of astronomical data sets, catalogues, and more general
concepts e.g. "quantity".

To describe 3D spectroscopic data we use Characterisation Data Model (Louys
et al. 2007). One of the most abstract data models developed by the DM WG,
it gives a physical insight to the dataset, i.e. describes where, how
extended and in which way the Observational or Simulated dataset can be
described in a multidimensional parameter space, having the following axes:
{\bf spatial}, {\bf time}, {\bf spectral}, {\bf observed} (e.g. flux,
polarimetric), as well as other arbitrary axes.  For every axis the three
characterisation properties are defined: {\bf coverage}, {\bf resolution},
and {\bf sampling}. There are four levels of details in the description of
the dataset: (1) {\bf location} or {\bf reference value} -- average position
of the data on a given parameter axis; (2) {\bf bounds}, providing a
bounding box; (3) {\bf support}, describing more precisely regions on a
parameter axis as a set of segments; {\bf map}, providing a detailed
sensitivity map.

Details about applying Characterisation Data Model to the 3D spectroscopic
datasets are given in Chilingarian et al. (2006). The algorithm for the
characterisation metadata computation is described there as well.

\section{3D Data Archives}
We have developed two data archives providing access to fully-reduced
``science-ready'' IFU and IFP datasets: ASPID-SR and GIRAFFE Archive. For
both archives the IVOA Simple Spectral Access (SSA, Tody et al. 2007)
interfaces are provided.

\subsection{ASPID-SR}
ASPID stands for the "Archive of Spectral, Photometric, and Interferometric
Data". The world largest collection of raw 3D spectroscopic observations of
galactic and extragalactic sources is provided. ASPID-SR (Chilingarian et
al. 2007) is a prototype of an archive of heterogeneous science ready data,
fed by ASPID, where we try to take full advantage of the IVOA
Characterisation Data Model. Multi-level Characterisation metadata is
provided for every dataset. The archive provides powerful metadata querying
mechanism (Zolotukhin et al. 2007) with access to every data model element,
vital for the efficient scientific usage of a complex informational system.
ASPID-SR is one of the reference implementation of the IVOA Characterisation
Data Model. The datasets are provided in several formats: stacked spectra,
regularly-gridded data cubes, and Euro3D FITS.

A high level of integration between the archive WEB interface and existing
VO tools is provided (see next section).

\subsection{Giraffe Archive}
GIRAFFE Archive (Royer et al., this conference) contents fully reduced data
obtained with the FLAMES/Giraffe spectrograph at ESO VLT. Data obtained with
all three observing modes of Giraffe: MEDUSA (multi-object spectroscopy),
IFU (multi-IFU spectroscopy), and ARGUS (single IFU) are provided. Raw
datasets are taken from the ESO archive after the end of their proprietary
period and reduced in an automatic way using the Giraffe data processing
pipeline. There is a possibility of accessing individual extracted 1D
spectra from the multi-object spectroscopic observations, as well as full
datasets in the Euro3D FITS format.

\section{Client Software}
Presently, there is a number of VO tools available that deal with images
(such as CDS Aladin) and 1-D spectra (ESA VOSpec, SpecView, SPLAT). However,
none of them is able to handle IFU datasets. 

In a framework of the VO Paris project (Simon et al. 2006) we have developed
VO Paris Euro3D Client specifically to deal with the datasets in the
Euro3D FITS format in a VO context. This tool interacts with CDS Aladin to
display position of the fibers (or slit) on the sky and display individual
extracted spectra in ESA VOSpec. Catalogue of positions of fibers (or slit
pixels) can be exported as VOTable.

VO Paris Euro3D Client is an open-source Java package, including basic
functions for the Euro3D FITS I/O and a graphical user interface.

Individual or co-added spectra can be extracted from the Euro3D FITS file
and exported as VOTable serialization of the IVOA Spectrum Data Model 1.0
(McDowell et al. 2007). All the interaction between applications is done
using PLASTIC (PLatform for AStronomical Tool InterConnection) -- a
prototype of the VO application messaging protocol.

Presently VO-Paris Euro3D Client is used as an integrated data visualising
software at ASPID-SR - Science-Ready Data Archive at the Special
Astrophysical Observatory of Russian Academy of Sciences.

\begin{figure*}
\includegraphics[width=17cm]{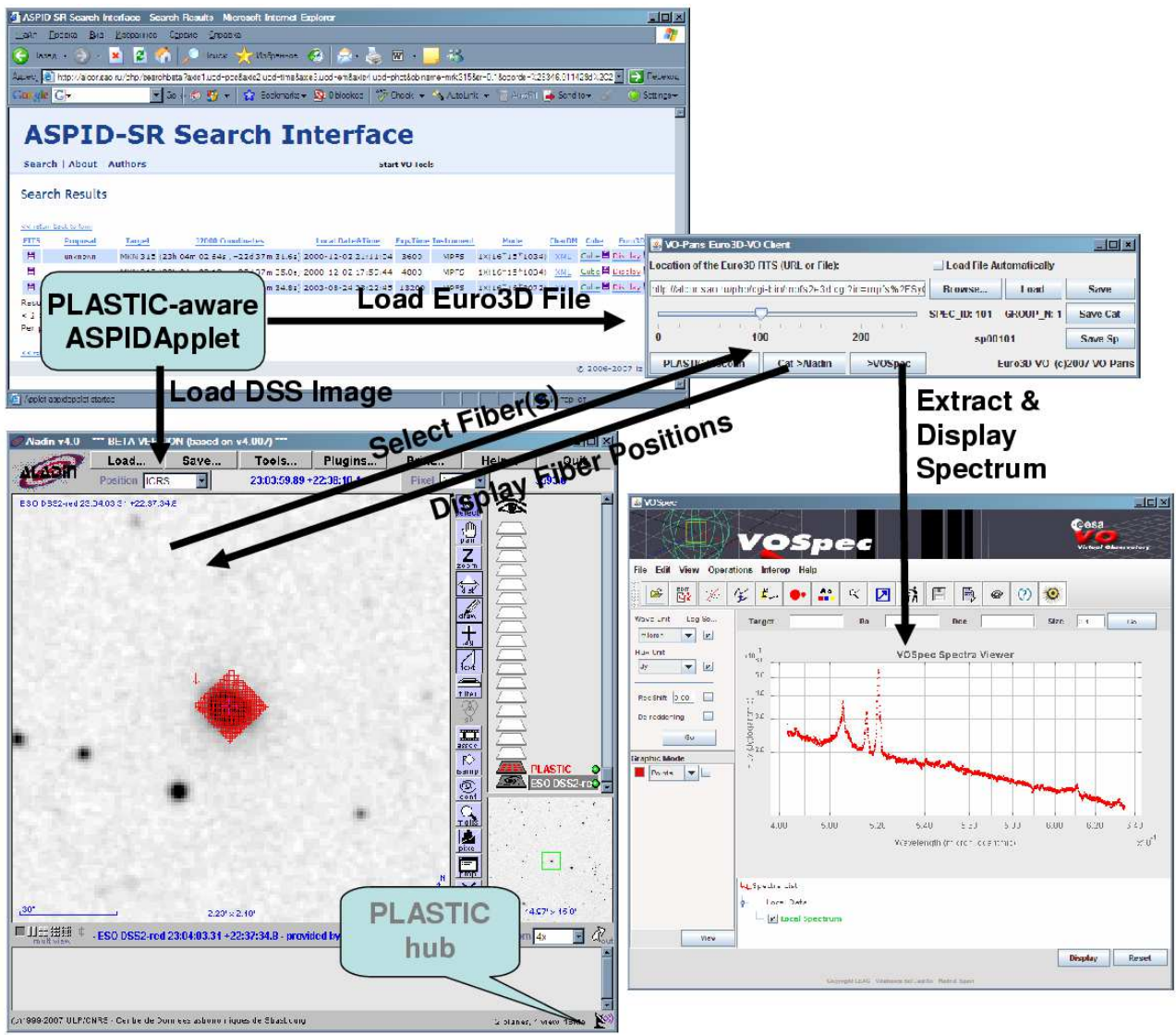}
\caption{Interaction between the ASPID-SR archive and PLASTIC-enabled VO tools. \label{figaspidsrplastic}}
\end{figure*}

In Fig.~\ref{figaspidsrplastic} we demonstrate how the interaction between 
VO client applications and the ASPID-SR archive interface is implemented. 
There are several stages:
\begin{enumerate}
\item Querying the characterisation metadata using WEB-interface (see Louys et al., 
this conference)
\item Light-weight Java applet is integrated into the HTML pages, containing query
response; it detects a PLASTIC hub, connects to it, and checks whether other
tools (Aladin, VOSpec, VO Paris Euro3D Client) are registered within it. If the 
applications are not detected, they will be started using JavaScript and Java WebStart.
\item As soon as all the used applications have been started and registered within the 
PLASTIC hub, a small script is sent to CDS Aladin to display the DSS2 image of the area,
corresponding to the position of the IFU spectrograph. At the same time, the IFU dataset
in the Euro3D FITS format is loaded into VO Paris Euro3D Client.
\item Positions of IFU fibers are sent from VO Paris Euro3D Client to CDS Aladin and 
overplotted on the DSS2 image.
\item User can interactively select either groups of fibers or individual ones using CDS Aladin.
An extracted spectrum (or co-added spectra of several fibers) is sent to ESA VOSpec using
PLASTIC by clicking on the corresponding button in the user interface of VO Paris Euro3D Client.
\end{enumerate}

\section{Summary}
In Chilingarian et al. (2006) we concluded that "all the necessary
infrastructural components exist for building VO-compliant archives of
science-ready 3D data and tools for dealing with them". Since that time
there was a substantial progress of VO standards and protocols. Now we are
able to provide access to first two such VO-compliant archives. This not
only a ``proof-of-concept'', but the services that can be used for real
scientific purposes. Another important conclusion that can be drawn is that
the present state of VO standards (including PLASTIC -- a prototype of the
VO application messaging protocol) is totally sufficient for dealing with
complex datasets in a VO framework without need to develop new client
applications for every particular kind of data.

\section*{Acknowledgments}
IC is grateful to ESAC and VO-Spain for providing financial support to
attend the workshop. Special thanks to John Taylor (Astrogrid) and Isa
Barbarisi (ESAC) for help with many technical points related to PLASTIC
implementation in VO Paris Euro3D Client.

\end{document}